\documentclass[twocolumn,noshowpacs,preprintnumbers,amsmath,amssymb]{revtex4}

\usepackage{verbatim}

\usepackage{graphicx}
\usepackage{dcolumn}
\usepackage{bm}
\usepackage{wrapfig}

\begin{document}

\title{Interaction of solid organic acids with carbon nanotube field effect transistors}
\author{Christian Klinke}
\email{klinke@chemie.uni-hamburg.de}
\altaffiliation{Present address: Institute of Physical Chemistry, Universtity of Hamburg, 20146 Hamburg, Germany.}
\affiliation{IBM T. J. Watson Research Center, 1101 Kitchawan Road, Yorktown Heights, NY 10598, USA}

\author{Ali Afzali}
\affiliation{IBM T. J. Watson Research Center, 1101 Kitchawan Road, Yorktown Heights, NY 10598, USA}

\author{Phaedon Avouris}
\affiliation{IBM T. J. Watson Research Center, 1101 Kitchawan Road, Yorktown Heights, NY 10598, USA}

\begin{abstract} 

\textit{A series of solid organic acids were used to p-dope carbon nanotubes. The extent of doping is shown to be dependent on the pKa value of the acids. Highly fluorinated carboxylic acids and sulfonic acids are very effective in shifting the threshold voltage and making carbon nanotube field effect transistors to be more p-type devices. Weaker acids like phosphonic or hydroxamic acids had less effect. The doping of the devices was accompanied by a reduction of the hysteresis in the transfer characteristics. In-solution doping survives standard fabrication processes and renders p-doped carbon nanotube field effect transistors with good transport characteristics.}

\end{abstract}

\maketitle

\section*{Introduction}

Carbon nanotubes (CNT) have been shown to be high-performance building blocks in electronic devices like field effect transistors (FET)~\cite{C01,C02}. They provide high carrier mobility~\cite{C03} and chemical sensitivity~\cite{C04}. The properties of such FETs depend very strongly on the metals used as leads~\cite{C05,C06} and the environmental conditions~\cite{C07,C08}. For example, due to absorption of oxygen at the metal contacts, intrinsic CNTs can form p-type devices~\cite{C07}. Recently, we reported chemical p-doping of CNT by treatment with a one-electron oxidizing agent to form air-stable p-type CNTFETs~\cite{C09}. Strong organic acids, like trifluoroacetic acid have been shown to be effective one-electron oxidants for electron rich organic compounds~\cite{C10}, while Dukovic et al. showed that carbon nanotubes can be protonated by strong acids~\cite{C11}. Thereby, the charge transfer depends on the chirality and accordingly on the bandgap of the nanotubes~\cite{C12}. CNTFETs are Schottky barrier transistors. Those barriers appear at the interface between the semiconducting nanotubes and the metallic leads. The workfunction alignment of the leads with the nanotube determines the extent of the Schottky barriers~\cite{C13,C14,C15}. The switching of the CNTFETs is based on the bending of the nanotube bands. The doping effect manifests mainly in the shifting of the threshold voltage to more positive (p-doping) or more negative (n-doping) values~\cite{C09,C16,C17}. In this communication we explore systematically the oxidation effect (p-doping) of various strong solid organic acids on CNTFETs and we show that the extent of doping depends strongly on the acidity of the organic acids. Direct doping of the devices with acids has the disadvantage of increasing the leakage current to the gate electrode. This could be prevented by doping the nanotubes already in solution followed by the fabrication of the devices with the functionalized CNTs.

\begin{figure}[!h]
\begin{center}
\includegraphics[width=0.45\textwidth]{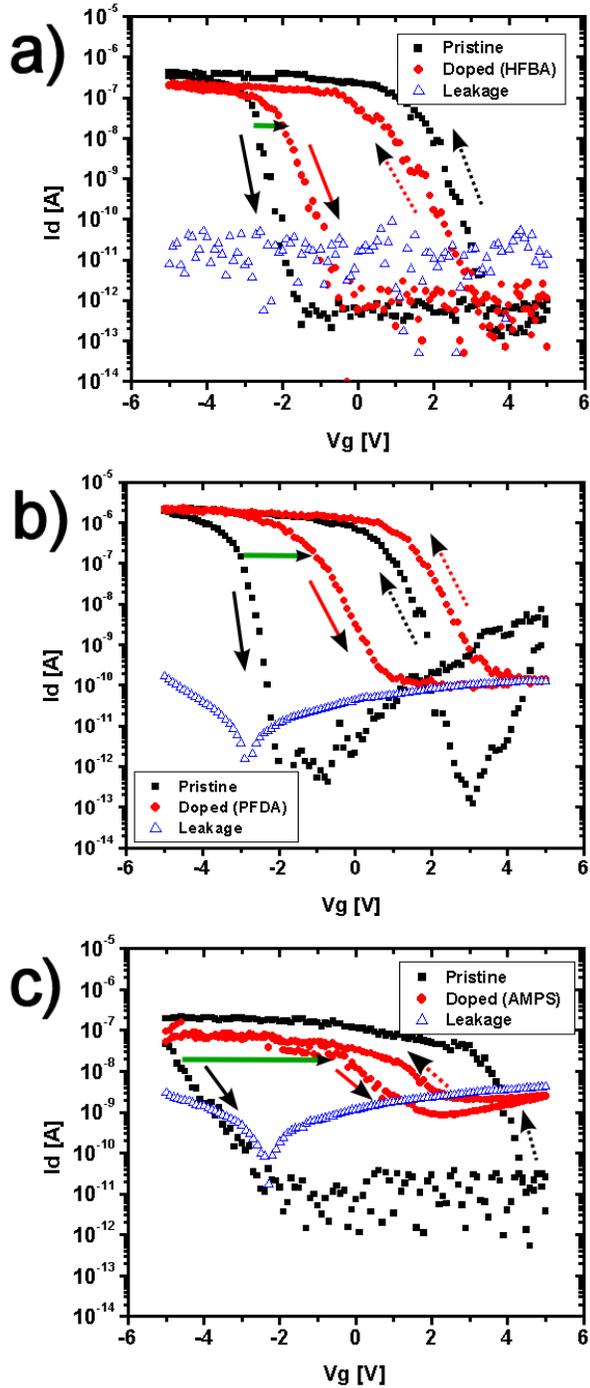}
\caption{\it Typical transfer characteristics (I$_{d}$ vs. V$_{g}$) of CNTFET devices with a source-drain voltage of V$_{ds}$ = -0.5~V on a 20~nm gate oxide, demonstrating the doping effect of the solid organic acids a) HFBA, b) PFDA, and c) AMPS. Arrows indicate the direction of voltage sweep (black - pristine device; red - doped device) and the threshold voltage shift (in green).}
\label{F01}
\end{center}
\end{figure}

\section*{Experimentals}

Carbon nanotube FETs were fabricated by the deposition of laser ablation carbon nanotubes from dispersion in dichloroethylene onto a silicon wafer with 20~nm thermally grown silicon dioxide. Field effect transistors with channel lengths of 400~nm were defined by e-beam lithography and successive e-beam deposition of a titanium adhesive layer (0.7~nm) and palladium (25~nm). The substrates with CNTFETs were then immersed in a 10~mM ethanolic solution of various acids, N-hydroxyheptafluorobutyramide (HFBA), perfluorododecanoic acid (PFDA), 2-methyl acrylamidopropane sulfonic acid (AMPS), aminobutyl phosphonic acid (ABPA), and hexadecyl phosphonic acid (HDPA) for 24~hours and then dried under a stream of nitrogen. The pKa value of those acids are (HFBA) $\sim$~4.5, (HDPA) $\sim$~2.6, (PFDA) $\sim$~1.0, and (AMPS) $\sim$~-2.0. The electric transport behavior of the CNTFET devices was characterized in a nitrogen-purged glove-box using an Agilent 4165C semiconductor Parameter Analyzer.

In order to evaluate the doping effect independently from the transport measurements we employed absorption spectroscopy on purified HiPCO nanotubes specifically for AMPS. We dried an ethanolic solution of nanotubes on a quartz plate. The absorption of this sample was then measured with a Perkin Elmer UV/Vis/NIR spectrometer Lambda~9. A 10~mM solution of AMPS, the strongest dopant, was then added dropwise to the sample.

\begin{figure}[!h]
\begin{center}
\includegraphics[width=0.45\textwidth]{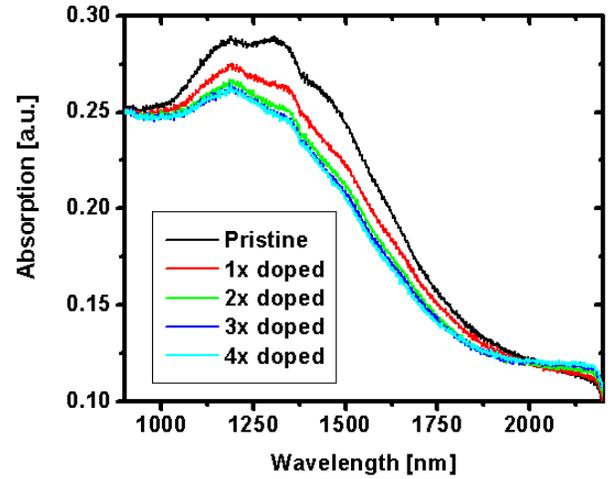}
\caption{\it Absorption spectroscopy of purified and dried nanotubes produced by the HiPCO method. The sample was stepwise doped with one droplet of a 10~mM AMPS solution. Arrows indicate the most prominent peaks.}
\label{F02}
\end{center}
\end{figure}

\section*{Results}

We measured the transfer characteristics of the CNTFETs before and after doping with the solid acids N-hydroxyheptafluorobutyramide (HFBA), perfluorododecanoic acid (PFDA), and 2-methyl acrylamidopropane sulfonic acid (AMPS) (Fig.~\ref{F01}). These compounds show increasing acidity in this sequence. In most cases the drive current I$_{on}$ (upper saturation current) decreases by less than one order of magnitude. The effect of doping, i.e. the threshold voltage shift~\cite{C18} increased with the strength of the acid (Tab.~\ref{T01}). The threshold voltage is defined by the upper point where the drain current I$_{d}$ deviates from exponential increase in the transfer characteristics. The threshold voltage was shifted to more positive values, i.e. the devices showed a more p-type behavior after doping. Additionally, in all cases the gate hysteresis decreased. Stronger acids decreased the hysteresis more. However, the leakage current (current to the gate) increased with the strength of the acid and the ON/OFF ratio decreased. The actual values are summarized in Tab.~\ref{T01}. The threshold shift vanished when the devices were measured in vacuum, and could be restored by exposing them to air - suggesting a cooperative effect between the acid and oxygen. 

\begin{table*}[tbp]
\begin{center}
\caption {\it Comparison table.}
\label{T01}
\vspace{0.5cm}
\begin{tabular}{|c|c|c|c|c|} \hline
\textbf{Acid} & \textbf{pKa} & \textbf{Threshold shift [V]} & \textbf{ON/OFF ratio (before/after)} & \textbf{Hysteresis (before/after) [V]}    \\ \hline
HFBA & 4.5~\cite{C23}  &  0.6  & 7.3E5 / 1.3E5 & 4.7 / 2.8  \\ \hline
HDPA & 2.0~\cite{C24}  &  2.1  & 1.3E5 / 3.4E2 & 6.3 / 2.5  \\ \hline
PFDA & 0.5~\cite{C25}  &  2.3  & 1.7E6 / 1.8E4 & 4.0 / 2.6  \\ \hline
AMPS & -2.0~\cite{C26} &  4.7  & 2.0E4 / 3.1E2 & 4.6 / 0.7  \\ \hline
ABPA & n/a             &  -1.6 & 6.6E5 / 2.0E5 & 4.2 / 4.3  \\ \hline
\end{tabular}
\end{center}
\end{table*}

The doping effect was confirmed independently by absorption measurements. First, the UV/Vis/NIR spectrum of purified HiPCO nanotubes dried on a quartz plate was measured. By adding dropwise a solution of 10~mM AMPS in ethanol, the intensity of the absorption peaks decreased substantially (Fig.~\ref{F02}). Larger wavelengths, corresponding to tubes with smaller bandgap, and thus tubes with larger diameters, were suppressed more effectively. The reduction of absorption peaks is typically attributed to electron depletion from or filling specific bands due to doping~\cite{C19}. The doping is more efficient for tubes with smaller bandgaps (larger wavelengths). AMPS itself is transparent in the range where the spectra were recorded. 

\begin{figure}[!h]
\begin{center}
\includegraphics[width=0.45\textwidth]{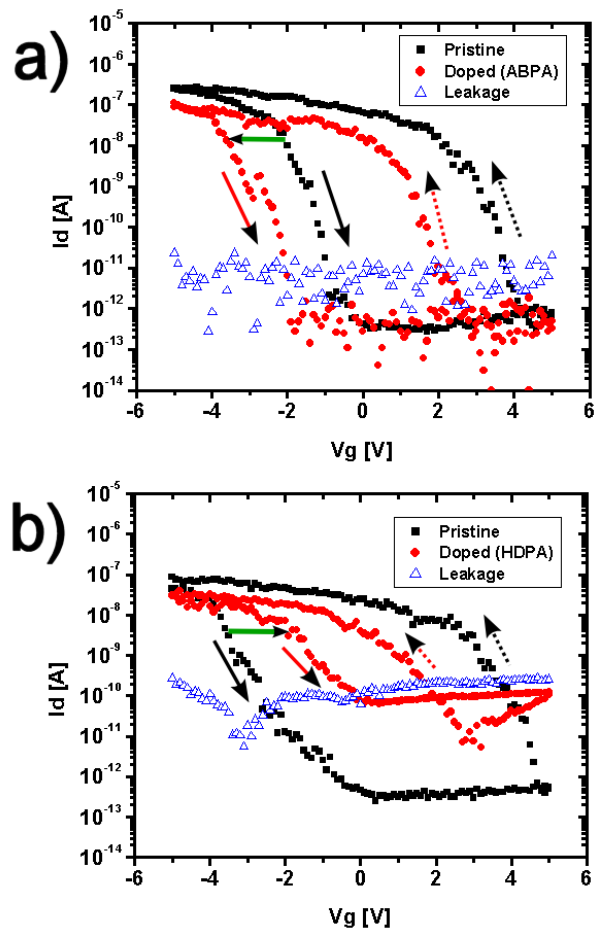}
\caption{\it Transport measurements with CNTFET devices demonstrating the doping effect of a) the bifunctional acid ABPA, and b) the solid acid HDPA. Arrows indicate the direction of voltage sweep (black - pristine device; red - doped device) and the threshold voltage shift (in green).}
\label{F03}
\end{center}
\end{figure}

In other experiments, bifunctional aminobutyl phosphonic acid (ABPA) containing both amine and phosphoric acid and hexadecyl phosphonic acid (HDPA) were used for doping experiments (Fig.~\ref{F03}). The doping effect of amino groups on carbon nanotubes have been studied earlier, and were shown to be strong n-dopants~\cite{C17}. Whereas HDPA with only a phosphonic acid group moderately p-dopes the CNT, treatment of CNTFETs with the bifunctional ABPA results in moderate n-doping. This experiment shows that the electron donating affinity of the amino group is higher than electron-accepting affinity of the phosphonic acid. In order to test the effect of the acid group we used HDPA. This molecule is similar to ABPA but without the amino group at one end. This molecule drives the device again to a more p-type characteristic.

\begin{figure}[!h]
\begin{center}
\includegraphics[width=0.45\textwidth]{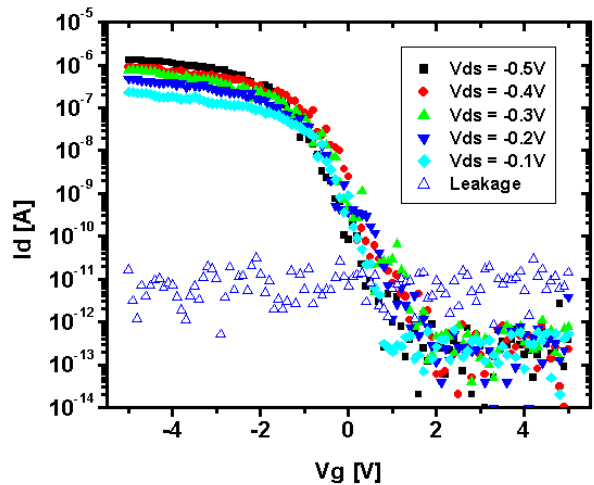}
\caption{\it Transport measurements with CNTFET devices demonstrating the effect of ex-situ doping with AMPS.}
\label{F04}
\end{center}
\end{figure}

Since the in situ doping increased the OFF-current (lower limit of I$_{d}$) significantly we doped the nanotubes with acids in solution and fabricated devices with these already doped tubes. For this we used again the strongest acid AMPS (Fig.~\ref{F04}). Under the consideration that our pristine nanotube devices usually switch at V$_{th}$ = -3~V ± 1~V, we conclude that the devices fabricated from solution-doped (ex situ) nanotubes switches at threshold voltages which are comparable to the in-situ doped nanotube, namely at -0.6~V for in situ doped vs. -1~V for ex situ doped devices.

\section*{Discussion}

The ability of highly acidic trifluoroacetic acid to oxidize electron-rich organic compounds to their radical cation has been reported by Eberson et al.~\cite{C10}. The oxidizing effect of trifluoroacetic acid is attributed to first protonation of electron-rich compounds followed by an electron transfer to an oxygen molecule. Since trifluoroactic acid has very high vapor pressure and is desorbed easily upon moderate heating, we were prompted to study the effect of higher molecular weight organic acids on oxidation (p-doping) of carbon nanotubes. We chose different classes of organic acids. PFDA with a pKa of about 0.5 and AMPS with a pKa of about -2.0 proved to be effective p-dopants and to shift the threshold voltage of CNTFETs to more positive voltages significantly as shown in Fig.~\ref{F01}. Milder organic acids like phosphonic acids (HDPA) and hydroxamic acids (HFBA) with a pKa value of about 2.5 and 4.5, respectively, were less effective and the shift to more positive threshold voltages were less pronounced. The contacts are not much affected during the doping of the whole device since the ON-current does not change much.

The effect is most likely due to protonation, followed by oxidation to radical cation in the presence of oxygen~\cite{C12,C20}. The presence of oxygen seems to be necessary for the stabilization of the protonation, since we find that the doping effect vanishes in vacuum and can be restored by exposure to air. The level of doping is determined by the strength of the acid i.e. the effectiveness of protonation of CNTs. Lower pKa value leads to stronger hole injection. 

The hysteresis in the transfer characteristic of CNTFETs might be due to organic contamination from the lithography process~\cite{C21}. The reduction of hysteresis is attributed to "cleaning" of the devices by the doping procedure (soaking the devices for 24~h in the corresponding acid solutions). This treatment might remove organic material around the carbon nanotubes. Similar cleaning effects were observed before~\cite{C22}.

We confirmed the doping effect independently by absorption measurements. Furthermore, we showed that the acids group of the molecules is responsible for the p-doping effect by comparing the molecules with an amino group containing acid. The known n-doping effect of amino groups~\cite{C17}, competes with the acid group. The final effect is an n-doping effect, whereas a similar molecule without the amino group does not show this effect, but a shift to more p-type behavior.

Direct doping of CNTFET devices leads also to higher OFF-currents due to higher leakage currents to the gate, and the subthreshold slopes become shallower. The higher leakage is due to the protonation of the oxide. We were able to circumvent this disadvantage by doping the nanotubes first in solution and then integrating the so-treated tubes into devices. This process restores the high performance of pristine nanotube devices while producing the desired threshold shift.

\section*{Conclusions}

To conclude, we used a series of solid organic acids in order to p-dope carbon nanotubes devices. The extent of doping was shown to be dependent on the pKa value of the acids. Highly fluorinated carboxylic acids and sulfonic acids were very effective in shifting the threshold voltage of carbon nanotube field effect transistors. Weaker acids like phosphonic or hydroxamic acids had less doping effect. The doping effect was also confirmed by UV/Vis/NIR absorption spectroscopy. The in-solution doping survives standard fabrication processes and renders p-doped CNTFETs with good transport characteristics. The influence of acids is not only interesting form a doping point of view, but acids are also present as contaminations in semiconductor processing from etching steps or from rinsing with solvents.

\section*{Acknowledgement}

We like to acknowledge gratefully the Alexander-von-Humboldt Foundation for financial support and Bruce Ek for technical assistance.

\clearpage

\end{document}